\documentclass[12pt]{article}

\usepackage{graphicx}

\title{  Nonperturbative QCD thermodynamics in the  external magnetic field}
\author{  M.A.Andreichikov and  Yu.A.Simonov \\
State Research
Center\\Institute of Theoretical and Experimental Physics, \\
Moscow, 117218 Russia}

\newcommand{\beq}{\begin{eqnarray}}
 \newcommand{\eeq}{\end{eqnarray}}
\newcommand{\be}{\begin{equation}}
 \newcommand{\ee}{\end{equation}}

\def\fun#1#2{\lower3.6pt\vbox{\baselineskip0pt\lineskip.9pt
\ialign{$\mathsurround=0pt#1\hfil ##\hfil$\crcr#2\crcr\sim\crcr}}}

\newcommand{{\SD}}{\rm SD}

\newcommand{{\Mc}}{\mathcal{M}}

\newcommand{\vep}{\mbox{\boldmath${\rm p}$}}

\newcommand{\veB}{\mbox{\boldmath${\rm B}$}}

\begin{document}
\maketitle
\begin{abstract}
 The thermodynamics  of quarks and  gluons   strongly depends on the vacuum
 colormagnetic field, which grows with the  temperature $T$, so that spatial
 string tension $\sigma_s ={\rm const}~ g^4 (T) T^2$. We  investigate below what
 happens when one imposes in addition constant magnetic field and  discover
 remarkable structure of the resulting thermodynamic potential.
 \end{abstract}

 \section{ Introduction}

The problem of quark gluon thermodynamics in  magnetic field is of the  high
interest in the modern physics, since heavy-ion experiments produce important
data on the properties of resulting hadron yields and hadron interactions,
which might be influenced by the strong magnetic  fields (MF) created during
the collision process \cite{1,2,3}. For a recent review on the  effects of MF
see \cite{4}.

 On the theoretical side the problem of MF in the quark gluon plasma (qgp) was
 studied in different aspects, e.g. in the NJL-type models \cite{5} and  in the
 holographic
 approach  \cite{6,7}. Within the nonperturbative QCD the theory of qgp in MF
 was developed in \cite{8,9,10}, where the general form of the thermodynamic
 potentials was found in MF with  zero or nonzero baryon density, summing over
 all Landau levels including LLL.

 In this approach the only nonperturbative interaction, which was taken into
 account, reduced to the inclusion of Polyakov lines in the  resulting expression for the  pressure.

 The resulting  expressions for magnetic  susceptibilities $\hat\chi_q(T)$,
 obtained in \cite{8}, were used in  \cite{9,10} to compare with the lattice
 data from \cite{11,12}, and  a reasonable  agreement was found for
 $\hat\chi_q(T)$with different $q=u,d,s$, in \cite{9,10} as  well as for the
 sum \cite{10}, however somewhat renormalised values  of effective quark masses
  were used.

  Recently in \cite{13,14} a new step in the  development of the np QCD
  thermodynamics was made, where the colormagnetic confinement (CC) was
  included in the dynamics of the qgp.  This interaction with the spatial
  string tension $\sigma_s$ grows with  temperature,$\sigma_s \sim g^4 T^2$ and
  is important in the whole region $T_c<T<10$ GeV. A concise  form of the final
  expression was found in \cite{13,14} in the case of an oscillatory type CC and
  an  approximate one in the realistic case of the linear CC. The resulting
  behavior both in the SU(3) case, found in \cite{15,16}, and in the qgp case
  \cite{13,14} agrees well with the corresponding lattice data.

  It is the purpose of the present paper to extend our previous  analysis of
  the qgp thermodynamics in MF, done in \cite{8,9,10}, including the dynamics
  of CC  with the explicit form of the magnetic screening mass $m_D,$ generated
  by CC.

  As will bee seen, we propose a simple generalization of the results
  \cite{8,9,10}, where the CC produces the mass  $M_D$,  entering the final
  expressions in the  combination $\sqrt{m^2_q  + m^2_D}  \equiv \bar M$
  instead of $m_q$. We check the limiting  cases and compare the result with
  lattice data.

  The paper is organized as follows. In the next section the general analysis
  of the MF effects in thermodynamics is  explained, in section 3  the magnetic
  susceptibility is  defined, in  section 4 the results are compared to the
  lattice data, and in the section 5 the summary and prospects are given.

  \section{General structure of the pressure with and without magnetic field}

  We start with the quark pressure of a given flavor as expressed via the 3d
  quark Green's function $S_3(s)$ in the  stochastic field of the colormagnetic
  confinement (CMC). From the path integral representation \cite{14,15} one obtains
  \be P^f_q = \frac{4N_c}{\sqrt{4\pi}} \int^\infty_0\frac{ds}{s^{3/2}}
  e^{-m_q^2 s} S_3 (s) \sum_{n=1,2,...} (-)^{(n+1)} e^{-\frac{n^2}{4T^2_s}}\cosh \frac{\mu n}{T} L^n\label{1}\ee
where $L=\exp \left(-\frac{V_1 (\infty, T)}{2T}\right)$ is the quark Polyakov
line, and $S_3(s)$ can be expanded in a series over eigenstates in the CMC on
the 2d minimal area  surface in 3d space. \be S_3 (s) = \frac{1}{\sqrt{\pi s}}
\sum_{\nu=0,1,...} \psi^2_\nu (0) e^{-m^2_\nu s},~~ m^2_0 = m^2_D = 4\sigma_s
(T).\label{2}\ee

As it was argued in \cite{14,15}, in the case of linear CMC one obtains for
$S_3(s)$ an approximate form \be S_3^{\rm lin} (s) \cong \frac{1}{(4\pi s)^{3/2}}
e^{-\frac{m^2_D s}{4}}\label{3}\ee and pressure can be written as \be
\frac{1}{T^4} P^f_q =\frac{N_c}{4\pi^2} \sum^\infty_{n=1} \frac{(-)^{n+1}}{n^4}
L^n \cosh \frac{\mu n}{T} \Phi_n (T), \label{4}\ee where $\Phi_n(T)$ is \be
\Phi_n (T) = \frac{8n^2\bar M^2}{T^2} K_2 \left( \frac{\bar M n}{T}\right), ~~
\bar M = \sqrt{m^2_q+ \frac{m^2_D}{4}}.\label{5}\ee

On another hand, using the  relation from \cite{8} $s=\frac{n\beta}{2\omega},
~~ \beta = 1/T$ and the representation \be \int^\infty_0 \omega d \omega
e^{-\left( \frac{m^2}{2\omega} + \frac{\omega}{2} \right) n\beta} = 2 m^2 K_2
(mn\beta), \label{6}\ee one obtains as in \cite{8} the pressure of the given
quark flavor $\left( P_q = \sum_f P_q^{(f)}\right)$.

\be P_q^{(f)} = \frac{N_c}{\sqrt{\pi}} \int \frac{d^3 p}{(2\pi)^3}
\sum^\infty_{n=1} (-)^{n+1} \sqrt{\frac{2}{n\beta}} \int^\infty_0
\frac{d\omega}{\sqrt{\omega}} e^{-\left( \frac{ m^2_q + \vep^2+\frac{ m^2_{D
 }}{4}}{2\omega} +\frac{\omega}{2} \right) n\beta}.\label{7}\ee

Let us now introduce  the magnetic field $\veB$ along the $z$ axis, so that our
system of quarks undergoes the influence of both CMC field and
(electro)magnetic field (MF) at the same time.

We consider the influence of the MF only, and write the corresponding equations
from \cite{8} \be P_q^f (B) = \frac{N_c | e_q B|T}{\pi^2} \sum_{n_\bot, \sigma}
\sum^\infty_{n=1} \frac{(-)^{n+1}}{n} L^n \cosh \left(\frac{\mu n}{T}\right)
\varepsilon^\sigma_{n_\bot} K_1 \left( \frac{n \varepsilon^\sigma_{n_\bot}}{T}
\right)\label{8}\ee where \be \varepsilon^\sigma_{n_\bot}= \sqrt{|e_qB|
(2n_\bot +1 - \bar\sigma) + m^2_q}, ~~\bar\sigma= \frac{e_q}{|e_q|}
\sigma_z.\label{9}\ee

It is interesting, that in the case $\mu=0$ one can replace in (\ref{7}) the
exponent (as was suggested in \cite{8}) \be \frac{m^2_q+\vep^2}{2\omega} +
\frac{\omega}{2} \to \frac{m^2_q + p^2_z + (2n_\bot +1 -\bar \sigma) |e_q
B|}{2\omega} +\frac{\omega}{2}\label{10}\ee and using the phase space in MF \be
V_3\frac{d^3p}{(2\pi)^3} \to \frac{dp_z}{2\pi} \frac{|e_q B|}{2\pi}
V_3,\label{11}\ee and the relation \be \int^\infty_0 d\omega
e^{-\left(\frac{\lambda^2}{2\omega}+\frac{\omega}{2}\right)\tau} = 2\lambda K_1
(\lambda \tau),\label{12}\ee one obtains the same  Eqs. (\ref{8}), (\ref{9}),
but with the replacement \be m^2_q \Rightarrow m^2_q + \frac{m^2_D}{4} = m^2_q
+ c\sigma_s,\label{13}\ee
where $c \simeq 1$ for $T \rightarrow \infty$.

Therefore in what follows we shall be using the Eq.(\ref{8}) with the
replacement in (\ref{9}), $m^2_q \to \bar M^2 = m^2_q + c \sigma_s$.

As was shown in \cite{8}, the form (\ref{8}) can be  summed up over $n_\bot,
\sigma$ to obtain the following result (we consider below for simplicity  only
the case $\mu=0)$
$$ P_q^{(f)} (B) = \frac{N_c | e_q B|T}{\pi^2}
\sum^\infty_{n=1} \frac{(-)^{n+1}}{n} L^n \left\{ \bar M K_1 \left( \frac{n\bar
M}{T} \right) + \right. $$ \be + \left. \frac{2T}{n}  \frac{|e_qB|+\bar
M^2}{|e_q B|} K_2 \left( \frac{n}{T} \sqrt{|e_q B | + \bar M^2}\right) -
\frac{n|e_q B|}{12T} K_0 \left( \frac{n}{T} \sqrt{\bar M^2 + |e_q
B|}\right)\right\}.\label{14}\ee

Note, that  the first term in (\ref{14}) appears from the lowest Landau levels
(LLL). For these levels it is known from analysis in \cite{17}, that the
asymptotic quark energy values do not depend on $eB$ and are equal to the
$\sqrt{\sigma}$ for small  quark   mass. This agrees with our values $\bar M =
\sqrt{m^2_q + \sigma_s} \approx \sqrt{c\sigma_s}$ and supports our  expression
(\ref{14}) at least in the high $e_q B$ limit, $e_qB \gg \bar M,  $ when the
second and the third term in (\ref{14}) tend to zero. Hence one obtains in the
limit $|e_qB| \gg \bar M,T$
\be P_q^{(f)} (B)_{|e_qB|\to \infty} = \frac{N_c |
e_q B|T}{\pi^2} \sum^\infty_{n=1} \frac{(-)^{n+1}}{n} L^n \bar M K_1
\left(\frac{n\bar M}{T}\right).\label{15}\ee

Note, that the factor $|e_qB|$ appears due to the phase space relation in MF,
Eq. (\ref{11}).

Now we turn to the limit of small MF, $|e_q B|\ll \bar M, T$. One obtains from
(\ref{14}) the  contribution of the second term only \be P_q^{(f)} (B\to 0)  =
\frac{2N_c  T^2 \bar M^2}{\pi^2} \sum^\infty_{n=1} \frac{(-)^{n+1}}{n} L^n
  K_2 \left(\frac{n\bar M}{T}\right) + O((e_q B)^2).\label{16}\ee

  One can compare (\ref{16}) with (\ref{4}), obtained in the case of zero MF,
  and insertion of (\ref{5}) in (\ref{4}) yields the same answer as in
  (\ref{16}).

  \section{Magnetic susceptibility of the quark matter}

  Using general expression for the quark pressure (\ref{14}), one can define a
  more convenient quantity, the magnetic susceptibilities $\hat \chi_q^{(n)},~ \hat \chi_q^{(2)}\equiv\hat
  \chi_q$,
  \be  P^f_q (B,T) - P^f_q (0,T) = \frac{\hat \chi_q}{2} (e_q B)^2 +
  O((e_qB)^4).\label{17}\ee

  To this end one expands the Mc Donald functions $K_n(\sqrt{n^2+b^2})$   entering in (\ref{14})in
  powers of $b$, following \cite{8}, and one obtains

  \be  P^f_q (B,T) - P^f_q (0,T) = \frac{  (e_q B)^2  N_c}{2\pi^2} \sum^\infty_{n=1}  (-)^{n+1}
  L^nf_n
   \label{18}\ee
   \be f_n = \sum^\infty_{k=0} \frac{(-)^k}{k!} \left( \frac{ne_q B}{2T\bar
   M}\right)^k K_k \left(\frac{n\bar M}{T}\right) \left[\frac{1}{(k+1) (k+2)} -
   \frac{1}{6} \right].\label{19}\ee

   As a consequence, one has for $\hat\chi_q$
   \be \hat\chi_q (T)   =
\frac{ N_c  }{3\pi^2} \sum^\infty_{n=1}  (-)^{n+1}  L^n
  K_0 \left(\frac{n\bar M}{T}\right).\label{20}\ee

  It is possible to sum up  the series over $n$ in (\ref{18}), when one exploits
  the representation
  \be  K_0 \left(\frac{n\bar M}{T}\right)= \frac12 \int^\infty_0 \frac{dx}{x} e^{-n \left(\frac{1}{x} + \frac{\bar M^2 x}{4T^2}\right)}
  .\label{21}\ee
  As a result one obtains for the quadratic magnetic susceptibility (ms)
 \be \hat\chi_q (T)   =
\frac{ N_c  }{3\pi^2} I_q, ~~ I_q =\frac12 \int^\infty_0 \frac{dx}{x} \frac{L
e^{-  \left(\frac{1}{x} + \frac{\bar M^2 x}{4T^2}\right)}}{1+Le^{-
\left(\frac{1}{x} + \frac{\bar M^2 x}{4T^2}\right)}}
  .\label{22}\ee

  The ms in (\ref{22}) is defined for a given quark flavor $q$, and  the total
  m.s. for the quark ensemble, e.g. for $2+1$ species of quarks can be written
  as
  \be \hat\chi_q (T)   = \sum_{q=u,d,s} \chi_q (T) \left(
  \frac{e_q}{e}\right)^2.\label{23}\ee

  \section{Results and discussions}
We have presented above the thermodynamic theory of quarks in the magnetic field, when quarks are affected also by Polyakov line interaction and the CC interaction, generalizing in this way our old results of \cite{8}-\cite{10}, where the CC part was absent.

We have included the CC interaction in the energy eigenvalues $\varepsilon_{n_{\perp}}^{\sigma}$, Eq.(\ref{9}) tentatively via the replacement (\ref{13}). This substitute can be corroborated in the case of lowest Landau levels with $\bar{\sigma} = 1,\ n_{\perp} = 0$, where magnetic field does not eneter, and the effective quark mass in subject to the CC interaction only. In the general case one can expect possible interference of $eB$ and CC terms, which can spoil the suggested replacement. 

To make this first analysis more realistic, we have checked the limits of small and large values of $eB$. In the first case we have shown the correct correspondence with the $eB=0$ result of \cite{13,14}, and in the second case of large $eB$, the leading linear in $eB$ term is just LLL term, which is not influenced by magnetic fields, except for the phase space redefinition. These results enable us to proceed with the analysis and comparisons of obtained equations with lattice data.

\begin{figure}
 \centering
  \includegraphics[width=0.8\linewidth]{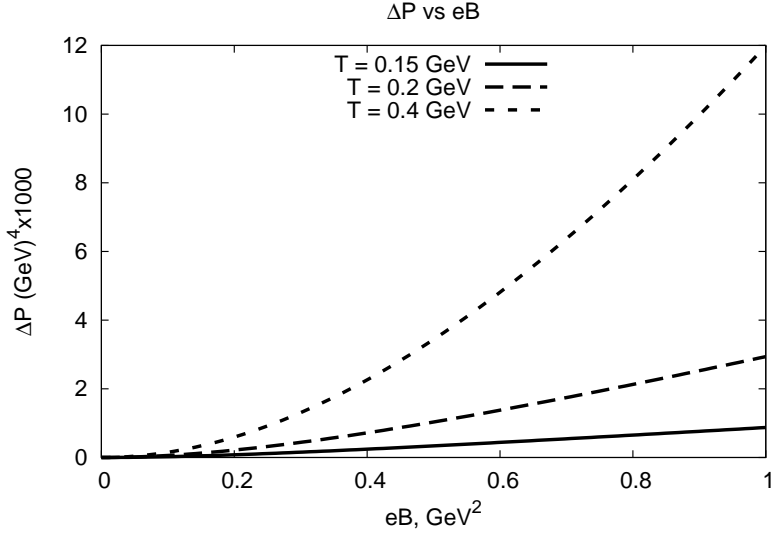}
  \caption{The dependence of $\Delta P(T,B)$ on $eB$ for fixed $T = 0.15,\ 0.2,\ 0.4\ GeV$. $\Delta P$ demonstrates quadratic behaviour $\sim(eB)^2$ for $eB < 0.5\ GeV^2$ and linear behaviour $\sim(eB)$ for $> 0.5\ GeV^2$ according to (\ref{14}). }
\label{fig1}
\end{figure}

\begin{figure}
 \centering
 \includegraphics[width=0.8\linewidth]{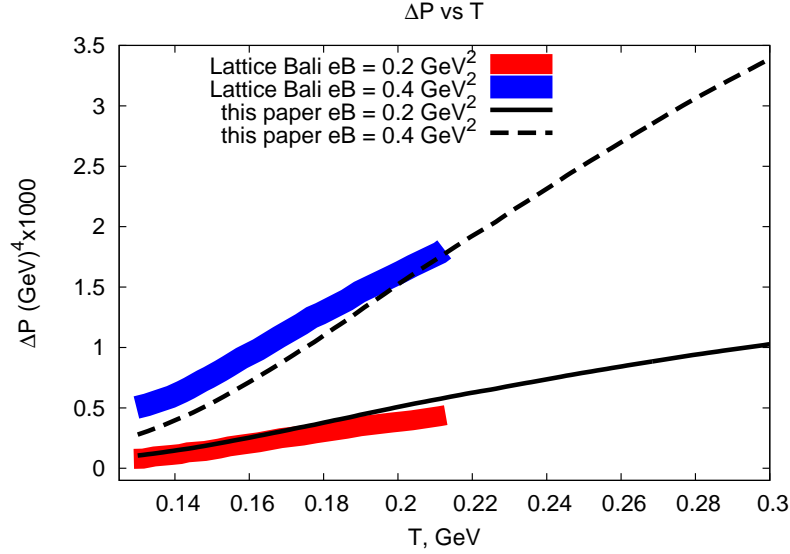}
 \caption{The dependence of $\Delta P(T,B)$ on $T$ for fixed values of $eB = 0.2,\ 0.4 \ GeV$ in comparison with lattice data from \cite{18}. For lattice results, the line thickness corresponds to the estimated error of the calculation.} 
\label{fig2}  
\end{figure}

  We present below an analysis of the MF influence on the quark  thermodynamics.  It is interesting, that the basic expression for the pressure $P_q^{(f)}$ in (\ref{14}) has the property, that at small $eB < (eB)_{crit}$, the dependence of $\Delta P(B,T)$ on $eB$ is quadratic with a good accuracy, according to Eq.(\ref{17}). At larger $eB$, $eB > (eB)_{crit}$, one has the linear dependence of $\Delta P(B,T)$ on $eB$, given in (\ref{15}). Fig.\ref{fig1} illustrates this behaviour for thee fixed temperatures $T = 0.15,\ 0.2,\ 0.4\ GeV$. One can see from (\ref{14}) and Fig.\ref{fig1} that $(eB)_{crit} > \bar{M}$, and actually is around $0.5 \ GeV^2$ for $T \simeq 0.2 \ GeV$. We have computed analytically the difference $\Delta
  P(T) = P(T, eB) - P(T,0)$ using Eq. (\ref{14}) and compare with the lattice  data from \cite{18}   for averaged $ u,d,s $ quark ensemble. One can see in Fig.\ref{fig2} the normalized pressure $\Delta P(B,T)$ for $T > 0.135 \ GeV$ and $eB = 0.2,\ 0.4\ GeV^2$ for the averaged quark ensemble of u,d,s quarks. In Fig.\ref{fig2} our calculations of $\Delta P(B,T)$ for $eB = 0.2,\ 0.4 \ GeV^2$ and $m_D^2 = 0.3\sigma_s$ using (\ref{14}) are compared with the lattice data from \cite{18}. We have used in (\ref{14}) the Polyakov line $L(T)$ obtained in \cite{19}. One can see a reasonable agreement within the accuracy of the lattice data, which supports the main structure of the theory used in the paper. Detailed analysis for larger intervals of $eB$ and $T$ and for specific quark flavours is possible within the approach and is planned for next publications.

\section*{Acknowledgement}
This work was done in the framework of the scientific project, supported by the Russian Scientific Fund, grant \#16-12-10414.

  \end{document}